\documentstyle[preprint,epsfig,aps]{revtex}
\begin{document}
\draft
 
\title{Near-Threshold $\eta$ Meson Production  
       in Proton--Proton Collisions}        
\author{J. Smyrski$^1$, 
        P. W\"{u}stner$^2$,
        J.T. Balewski$^{3,4}$,
        A. Budzanowski$^3$,
        H. Dombrowski$^6$,
        D.~Grzonka$^5$,
        L. Jarczyk$^1$,
        M. Jochmann$^2$,
        A. Khoukaz$^6$,
        K. Kilian$^5$,
        P. Kowina$^{5,7}$,
        M.~K\"{o}hler$^2$,
        T. Lister$^6$,
        P. Moskal$^{1}$,
        W. Oelert$^5$,
        C. Quentmeier$^6$,
        R. Santo$^6$,
        G. Schepers$^{5, 6}$,
        U. Seddik$^8$,
        T. Sefzick$^5$,
        S. Sewerin$^5$, 
        A. Strza\l kowski$^1$,
        M. Wolke$^5$}   
\address{$^1$ Institute of Physics, Jagellonian University, PL-30-059 Cracow, Poland}
\address{$^2$ ZEL,  Forschungszentrum J\"{u}lich, D-52425 J\"{u}lich,  Germany}
\address{$^3$ Institute of Nuclear Physics, PL-31-342 Cracow, Poland}
\address{$^4$ IUCF, Bloomington, Indiana, IN 47405, USA}
\address{$^5$ IKP, Forschungszentrum J\"{u}lich, D-52425 J\"{u}lich, Germany}
\address{$^6$ IKP, Westf\"{a}lische Wilhelms--Universit\"{a}t, D-48149 M\"{u}nster, Germany}
\address{$^7$ Institute of Physics, Silesian University, PL-40-007 Katowice, Poland} 
\address{$^8$ NRC, Atomic Energy Authority, 13759 Cairo, Egypt}

\date{\today}
\maketitle
\begin{abstract}
The production of $\eta$ mesons has been measured in the proton-proton 
interaction close to the reaction threshold using the COSY-11 internal 
facility at the cooler synchrotron COSY.  Total cross sections were determined 
for eight different excess energies ($\epsilon$) in the range from 
$\epsilon$~=~0.5~MeV to $\epsilon$~=~5.4~MeV.
The energy dependence of the total cross section is well described
by the available phase--space volume weighted by FSI factors for the 
proton--proton and proton--$\eta$ pairs. \\
\end{abstract}


\section{Introduction}
Over the last few years, creation of mesons near threshold 
in the elementary
nucleon--nucleon collision has become an important field for studies
of meson production mechanisms as well as of meson--nucleon interactions.
Measurements at the new generation of medium energy proton accelerators,
storage rings with phase-space cooling of the beam as the IUCF--ring,
CELSIUS and COSY, delivered high precision values of cross sections
for the production of  various mesons in the mass region up to 1~GeV/c$^{2}$. 
The experimental information gained so far 
is consistent with approximately constant production matrix elements when the 
final state interaction (FSI) is factored out. 
The pion production cross sections are described very precisely including
only the proton--proton FSI, since the pion--proton interaction is comparatively
weak close to threshold. Contrary, in the $\eta$ meson production in 
proton--proton collisions the $\eta$--proton FSI can essentially influence the 
energy dependence of the total cross section.
Effects of $\eta$--proton FSI have been seen in Dalitz plots from investigations 
at CELSIUS \cite{Cale96}. Inclusive measurements at CELSIUS \cite{Cale99} 
indicate additional contributions from partial waves of higher order than 
s--wave even at an excess energy as low as 36 MeV, 
which again change the energy dependence of the cross section 
according to their relative strength.\\[-0.9cm]

\section{Experiment}
Existing data on the $pp \rightarrow pp\eta$ reaction near threshold, 
originating from measurements at SATURNE
using the spectrometers SPES--3 \cite{Boud95} and PINOT \cite{Chia94}
and at CELSIUS \cite{Cale96} with the PROMICE/WASA detection system, still leave  
enough freedom for interpreting the energy dependence of the cross sections.\\
Therefore, further data of the $pp$ induced $\eta$ production very close
to threshold were needed.  
Measurements were performed at the COoler SYnchrotron COSY\cite{COSY} in 
J\"ulich  with the use of the COSY--11 facility, shown schematically in 
Fig.\ \ref{eta_topview},
in the range of excess energies below $\epsilon$~=~6~MeV.\\
The COSY-11 facility, described in detail in Ref.\ \cite{Brau96},
uses an internal hydrogen cluster target \cite{Domb97},
installed in front of a COSY accelerator dipole magnet.
Due to their lower momenta, the two outgoing protons of the reaction $pp \to pp \eta$ 
are separated from
the beam in the magnetic field of the C--shaped dipole and are diverted towards 
the direction of the centre of the synchrotron into the COSY-11 detector
arrangement. Their trajectories are measured by means of hits in  
a set of two drift chambers 
(marked D1 and D2 in Fig.\ \ref{eta_topview}), allowing the momentum to be 
determined by ray tracing back through the precisely known magnetic field 
to the target position. Identification of particles (here protons) is performed 
by additionally measuring the time-of-flight over a distance of 
$\approx$~9.4~m between start and stop 
scintillator hodoscopes (S1 and S3).
The uncharged $\eta$ mesons are not registered exclusively but are 
identified using the missing mass method. \\
Measurements were performed with the beam momentum varied continuously
in the range from 9.6~MeV/c below to 20.4~MeV/c above the
threshold momentum which is equal to 1981.6~MeV/c.
For the data analysis, this range was grouped into 2~MeV/c intervals.
In the following, the central momenta for these intervals 
and not their limits are quoted. \\
 \\
For different excess energies examples of missing mass distributions are shown 
in \mbox{Fig.\ \ref{eta_mismas},} each of them being dominated by  
a clear peak due to the $\eta$ meson production, except for the case where the 
beam momentum is below the reaction threshold. The 
$\eta$ missing mass distribution broadens with increasing excess energy, which 
is a kinematical effect. The square of the missing mass (MM) is determined by 
the square of the four-momentum vector evaluated when
subtracting the sum of the two exit proton four momentum vectors 
\mbox{({\bf{P$_1$~+~P$_2$}})} from the one of the proton--proton entrance channel 
({\bf{P$_0$}}):
\begin{equation}
MM^2 = ({\bf{P_0}} - ({\bf{P_1}} + {\bf{P_2}}))^2.
\end{equation}
For the limit of $\epsilon \ll m_{\eta}$ it can be shown 
\cite{Jurek_99} that the                
resolution of the square of the missing mass is proportional to the 
experimental momentum resolution of the two protons measured (which is 
supposed to be constant for all excess energies in the range discussed here) 
times $\sqrt {\epsilon}$:
\begin{equation}
\Delta \left(MM^2\right) = a \cdot \sqrt{\epsilon} .
\label{mm2res}
\end{equation} 
Fitting the width distribution ($\Delta \left(MM^2\right)$) of the $\eta$ missing mass peak
as function of the excess energy
in the present experiment at proton beam momenta around 2.0~GeV/c,
as shown in Fig.\ \ref{mismas_width}, 
a value of $a_{\eta}~=~(390~\pm~20)~\sqrt{MeV^{3}}/c^{4}$ was extracted.
 
%

\section{Data Evaluation}
 
The number of events corresponding to the $\eta$ meson production
was derived from the missing mass spectra.
The background underneath the $\eta$ peak, being due to the production of two to
four pions, was determined using measurements below threshold. 
For this reason, a smoothed background measured below threshold
has been shifted according to the kinematical limit of the missing mass spectra, 
shown as a dotted line in Fig.\ \ref{eta_mismas},
and was scaled according to the ratio of luminosities for the beam momenta above and 
below threshold. The background is that low 
(peak to  background ratio $\ge$ 40/1)  that the approximation of the 
smoothed background shown by the dashed line in Fig.~\ref{eta_mismas}
can hardly be seen.\\

Due to the rapid variation of near-threshold cross sections as function of beam 
momentum, a high precision knowledge of the absolute value of the beam momentum 
is extremely crucial for the present measurements. The present ''nominal'' 
beam momenta in the range around 2 GeV/c,
calculated from the synchrotron frequency and the beam orbit length,
are known at COSY with an accuracy of 
$\frac{\Delta p}{p}~=~10^{-3}$~\cite{COSY}. 
The corresponding uncertainty of the total cross section amounts to values as
large as $\frac{\Delta\sigma_{TOT}}{\sigma_{TOT}}~\approx~\pm~50~\%$
at $\varepsilon = 2$~MeV. \\

When evaluating the missing mass spectra with the nominal COSY beam momenta 
the average of the $\eta$ meson missing mass is shifted by about
+~0.66~MeV/c$^{2}$ compared to the $\eta$ meson mass known from 
literature~\cite{Euro98}, as indicated in Fig.\ \ref{eta_mismas} by an arrow. 
This discrepancy might partly be due to a systematic uncertainty in the 
detection system as incorrect assumptions of the magnetic fringe field or of 
the positions of the drift chambers \mbox{($\le\mid 0.28$~MeV/c$^2 \mid$)}~\cite{Jul97}.
However, the corresponding correction of the beam momentum of
$\Delta p= (-1.88\pm~0.80)$~MeV/c is in accordance with the 
$\frac{\Delta p}{p}~=~10^{-3}$ uncertainty of the nominal beam momentum. 

The relative uncertainty of the corrected
beam momentum of  $\delta p/p=0.4\cdot10^{-3}$ is by a factor
of two and a half smaller than the uncertainty of the beam momentum determined
from the beam orbit length and the frequency of the synchrotron.\\

In the experiment proton--proton elastic scattering was measured simultaneously.
The luminosity was determined by comparing differential counting rates with data 
obtained by the EDDA collaboration \cite{Albe97}.\\

\section{Determination of the Excess Energy}

Above, the best value for the true beam momenta at the eight different momentum
intervals was determined by shifting the extracted $\eta$ meson mass to its 
value known from the literature~\cite{Euro98}. 
 
In the following we present a second method for a determination of the beam 
momenta, where the measured dependence of the $pp \rightarrow pp\eta$ counting 
rate on the beam momentum can be used to evaluate the beam momentum
with high precision.  
The applied method is analogous to the one used by the COSY-11
collaboration for the beam energy determination in measurements
of the $pp \rightarrow p K^{+}\Lambda$ reaction \cite{Bale98}
and is largely independent of systematical uncertainties due to the magnetic 
fringe field or drift chamber positions. 
However, it assumes a phase space dependence modified by final state 
interactions of the total cross section close to threshold.
The measured yield of the $pp \rightarrow pp \eta$ events ($N$) normalized to 
the integrated luminosity~($L$) is extrapolated as a function of the excess 
energy towards zero. The corresponding offset of the excess 
energy was used to correct the nominal value of the excess energy:
\begin{equation}
N/L=C \cdot A(\varepsilon-\Delta\varepsilon) \cdot
\sigma(\varepsilon-\Delta\varepsilon),
\label{energy_calibration}
\end{equation}
where $C$ is a normalization factor, $A$ is the acceptance of the detection 
system, $\sigma$ is the total cross section and $\Delta\varepsilon$ is the
searched correction of the nominal excess energy. The values of $C$ and 
$\Delta\varepsilon$ are adjusted by the fitting procedure.
The efficiency $A$ is calculated using Monte Carlo simulations of the 
experiment assuming a uniform phase-space distribution of reaction products
modified by the $pp$ FSI. The influence of the $\eta p $ FSI  on the
acceptance is negligible.

The geometrical acceptance of the COSY-11 detection system is limited 
especially in the vertical direction due to the narrow opening of the 
dipole gap.
The calculated efficiency includes also the inefficiency of detecting
two close tracks due to limited double track resolution in the drift
chambers equal to 3~mm. 

The overall efficiency decreases from 31~\% to 4.4~\% in the 
range between the lowest and the highest excess energy measured. The energy 
dependence of the total cross section $\sigma$ was assumed to be determined by 
the phase-space volume weighted by the proton--proton FSI factor, which
was calculated as the squared absolute value of the complex amplitude
of the proton--proton scattering amplitude in the effective range
approximation with included Coulomb barrier penetration factor \cite{Mort68}.\\

As shown in Fig.\ \ref{eta_energy}, the experimental data are well 
described by the applied function.
The obtained correction to the excess energy is
$\Delta\varepsilon= (-0.66\pm0.27$)~MeV, which means that the real excess energy
is by that value smaller than the nominal one.
The indicated error contains contributions due to final statistics
of the data ($\pm 0.06$~MeV), due to the uncertainty of the detector acceptance 
($\pm 0.09$~MeV), and due to the uncertainty of the mass of the 
$\eta$ meson~(547.30~$\pm$~0.12) MeV/$\mbox{c}^2$\cite{Euro98}, which influences 
the present result via the threshold energy.
The correction coincidences with the value found from the shift of the
missing mass peak, the latter, as a model--free measure of the true 
excess energy, is used in the following.  
 
\section{Total Cross Sections}
The values of the total cross sections are given in Table\ \ref{cross_eta}
and are depicted in Fig.~\ref{eta_sigma_neu}. The indicated vertical error bars 
denote the statistical uncertainty only.
The overall systematical error amounts to 15~\%, where 10~\% originate from
the determination of the detection efficiency and 5~\% from the luminosity
determination. Data from measurements at SATURNE with the spectrometers 
SPES--3~\cite{Boud95} and PINOT~\cite{Chia94} and at CELSIUS \cite{Cale96} 
using the PROMICE/WASA system are added in Fig.\ \ref{eta_sigma_neu} to the 
present results.  The data are consistent and determine rather precisely 
the excitation function in the full range of the considered excess energies 
up to 40~MeV.

In order to describe the shape of the energy dependence of the cross
sections  
one can assume that it is dominated by the available phase-space
weighted by the $pp\eta$ FSI factor. 
The $pp\eta$ FSI can be factorized into $pp$ ($f_{pp}$) and $p\eta$ 
($f_{p\eta}$) factors and 
integrated over the available phase-space volume {\it{$\rho_{3}$}}:
\begin{equation}
\sigma(\varepsilon) \sim \int f_{pp}(q_{pp}) \cdot f_{p\eta}(q_{p_{1}\eta})
\cdot f_{p\eta}(q_{p_{2}\eta}) d\rho_{3},
\label{fsi_ppeta}
\end{equation}
where $q_{pp}$ is the relative momentum of two protons and $q_{p_{1}\eta}$ and 
$q_{p_{2}\eta}$ are the relative momenta of the $\eta$ meson with respect to first 
and second proton, respectively. The enhancement factors for the 
$\eta$--proton FSI were calculated in the effective range approximation with
the complex $\eta$--proton scattering length $a_{\eta p}=0.717+i0.263$ taken 
from Ref.\ \cite{Tshl98} and the complex $\eta$--proton effective range 
parameter $r_{\eta p}=-1.50-i0.24$ from Ref.\ \cite{Gree97}.
The calculations shown by the solid line in Fig.\ \ref{eta_sigma_neu} describe
the experimental data in the whole range of excess energy. Omission of the 
$\eta$--proton FSI leads to discrepancies with the data as shown by the 
dashed curve. At the same time, a calculation neglecting the proton--proton 
Coulomb interaction (dotted curve) fails to reproduce the energy dependence 
of the data within the limits of the relative uncertainty in the true beam 
energy.

\section{Conclusions}
The total cross section of the reaction $pp \rightarrow pp \eta$ 
was measured close to the kinematical threshold. The present results together 
with other available data determine the energy dependence of the cross section
in a wide excess energy range above threshold. This dependence is well 
reproduced by the phase-space integral, weighted by the full $pp \eta$ 
final--state interaction. In particular, inclusion of the $\eta$--proton FSI
as well as the proton--proton Coulomb interaction is essential for
the description of the data.\\
 \\
{\bf{Acknowledgements}}\\
We appreciate the work provided 
by the COSY operating team and thank them for the good cooperation and
for delivering the excellent proton beam. The research project was supported by
the BMBF (06MS881I), the Polish Committee for Scientific Research 
(2P03B-047-13), and the Bilateral Cooperation between Germany and Poland 
represented by the Internationales B\"{u}ro DLR for the BMBF (PL-N-108-95). 
The collaboration partners from the Westf\"{a}lische Wilhelms-University 
of M\"{u}nster and the Jagellonian University of Cracow 
appreciate the support provided by the FFE-grant (41266606 and 41266654,
respectively) from the Forschungszentrum J\"{u}lich. One of the authors (P.M.)
acknowledges the hospitality and financial support from the Forschungszentrum
J\"ulich and the Foundation for Polish Science.

\begin{table}
\caption[cross_eta]{
Total cross sections for the $pp \rightarrow pp\eta$ reaction
}
\label{cross_eta}
\begin{tabular}[]{ccc}
Beam momentum&Excess energy&Total cross section\\
(GeV/c)& (MeV)&($\mu$b)\\
\hline
1.98315& 0.54 &0.022 $\pm$~0.004  \\
1.98515& 1.24 &0.121 $\pm$~0.012  \\
1.98715& 1.94 &0.27  $\pm$~0.02   \\
1.98915& 2.64 &0.37  $\pm$~0.03   \\
1.99115& 3.34 &0.49  $\pm$~0.04   \\
1.99315& 4.04 &0.70  $\pm$~0.05   \\
1.99515& 4.73 &0.68  $\pm$~0.06   \\
1.99715& 5.43 &0.89  $\pm$~0.07   \\
\end{tabular}
 \end{table}
 The uncertainties of the beam momenta and the excess energies are:
 
 $\pm$~0.00080~GeV/c and $\pm$~0.28~MeV, respectively.
 
 The listed errors of the total cross sections include statistical uncertainties
 only, 
 
 the additional systematic error amounts to $\pm$~15$\%$.

\newlength{\ts}
\setlength{\ts}{\textwidth}
\addtolength{\ts}{-\columnsep}
\setlength{\ts}{0.5\ts}

\begin{figure}[t]
\epsfxsize=\ts
\epsffile{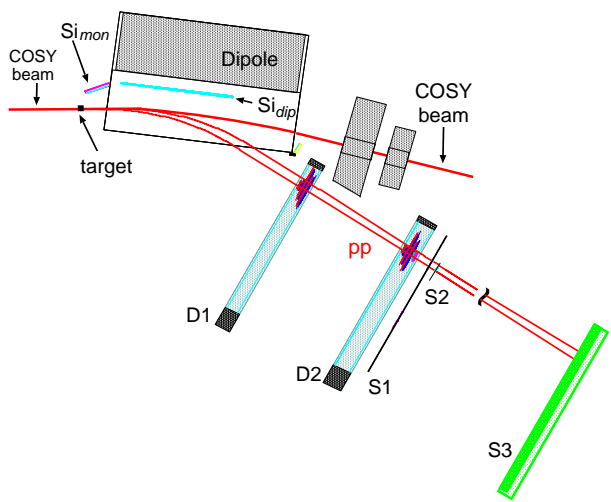}
\vspace{1.0cm}
\caption{Schematic view of the COSY-11 facility.
The particle trajectories are measured by means of hits in two sets of 
drift chambers D1 and D2. 
The scintillation hodoscopes S1 and S2 are used as start detectors 
and S3 as the corresponding stop detector for time of flight measurements.}
\label{eta_topview}
\end{figure}

\newpage
 
\begin{figure}[t]
\epsfxsize=\ts
\epsffile{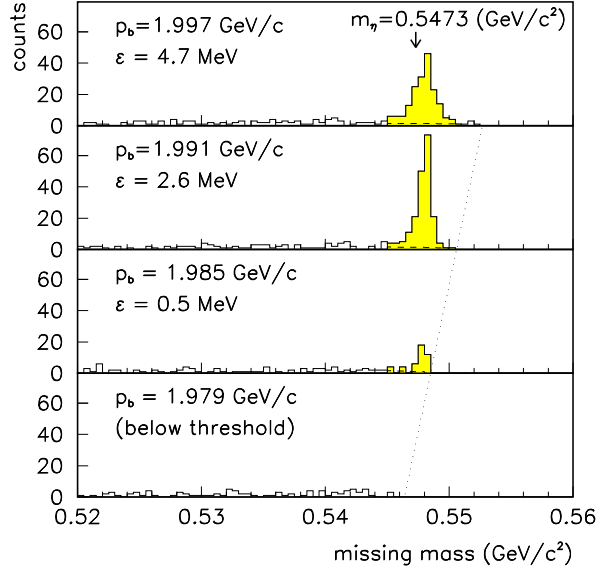}
\caption{Missing mass spectra with respect to the proton--proton system 
measured for three 
different momenta above and one momentum below threshold. The dotted line 
indicates the kinematical limit of the missing mass distribution. 
The dashed line underneath 
the $\eta$ peaks represents the background determined from measurements below 
threshold.}
\label{eta_mismas}
\end{figure}

 \begin{figure}[t]
\epsfxsize=\ts
\epsffile{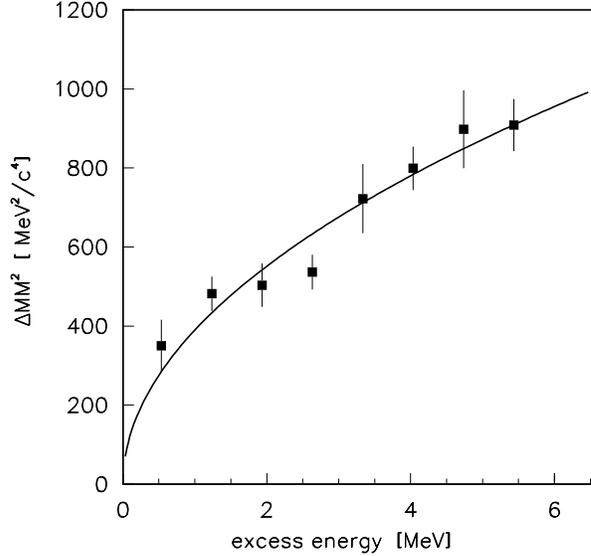}
\caption{Width of the $\eta$ missing mass peak as a function of the excess
energy. Experimental data points represent the width extracted from the central
momenta of the momentum bins applied, the solid line corresponds to a fit 
using eq.~(\ref{mm2res}) with the parameter 
$a_{\eta}~=~390~\sqrt{MeV^{3}}/c^{4}$.}
\label{mismas_width}
\end{figure}

\newpage

 \begin{figure}[t]
\epsfxsize=\ts
\epsffile{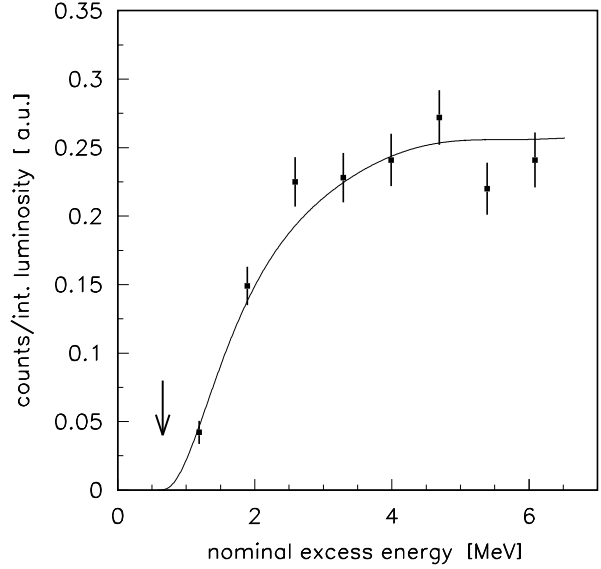}
\caption{Determination of the absolute value of the beam momentum by 
extrapolating the $pp \rightarrow pp\eta$ counting rate towards the threshold. 
The curve is given by eq.~\ref{energy_calibration}, 
the fit results in $\chi^2 = 1.5$ per degree of freedom.
The arrow corresponds to the resulting correction of the excess energy of 
$\varepsilon= (-0.66\pm0.27)$~MeV.}
\label{eta_energy}
\end{figure}

\begin{figure}[t]
\center
\epsfxsize=\ts
\epsffile{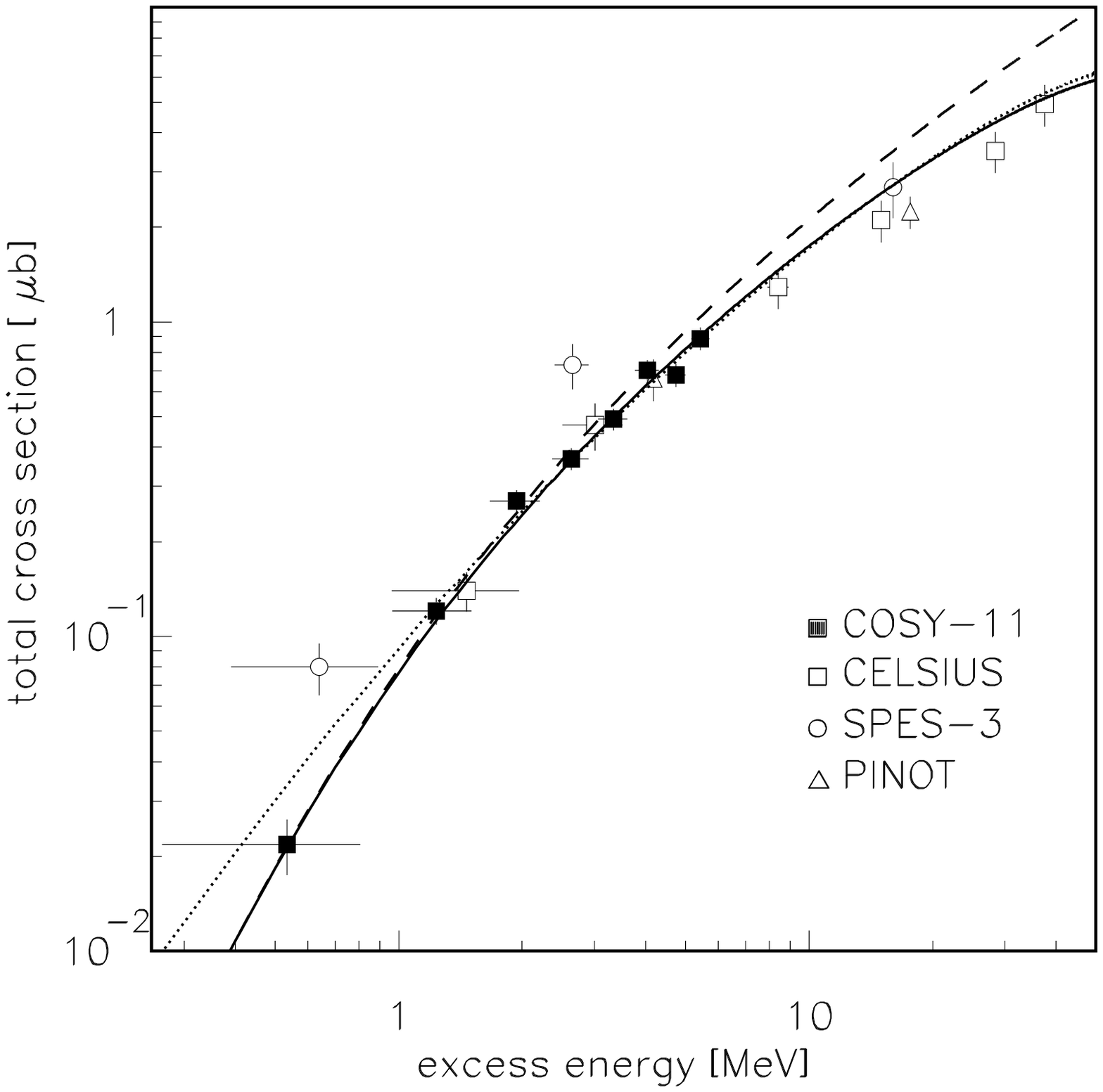}
\caption{Total cross sections for the reaction $pp \rightarrow pp \eta$ near 
threshold.\protect\\
Vertical error bars denote the statistical errors and the horizontal 
error bars show the systematical error of 0.28~MeV.
The relative excess energy uncertainty is 0.01~MeV (see Ref.~12), 
i.e. below the size of the symbols.\protect\\
Solid line: Phase-space distribution weighted by the 
proton--proton FSI \protect\\
$~~~~~~~~~~~~~~~$and Coulomb interaction 
plus proton--$\eta$ FSI,\protect\\
Dashed line: omission of proton--$\eta$ FSI from the solid line,\protect\\
Dotted line: omission of proton--proton Coulomb interaction from the solid line.}
\label{eta_sigma_neu}
\end{figure}

 \end{document}